\documentclass[journal=jpccck,manuscript=article]{achemso}

\author{Juan I. Climente}
\author{Jose L. Movilla}
\affiliation[Universitat Jaume I]
{Departament de Quimica Fisica i Analitica, Universitat Jaume I,
E-12080 Castello, Spain}
\author{Guido Goldoni}
\email{guido.goldoni@unimore.it}
\affiliation[University of Modena and Reggio Emilia]
{Department of Physics, University of Modena and Reggio Emilia, IT-41100 Modena, Italy}
\alsoaffiliation[CNR-S3]{Research Center S3, CNR-Institute of Nanoscience, IT-41100 Modena, Italy}
\author{Josep Planelles}
\email{josep.planelles@uji.es}
\affiliation[Universitat Jaume I]
{Departament de Quimica Fisica i Analitica, Universitat Jaume I,
E-12080 Castello, Spain}

\title[Excitons in nano-hybrids]
{The excitonic resonance in semiconductor-metal nano-hybrids}

\begin{document}

\begin{abstract}
We use a configuration interaction approach within the envelope function approximation to study the nature of the excitonic resonance in nano-hybrids, composite nanoparticles (NPs) combining a semiconducting and a metallic segment in contact. With reference to recent experimental reports, we specifically study CdS-based nanorods with metallic NPs deposited at the tips (matchstick) or metallic coatings (core-shell). The excitonic states are computed taking into account both the renormalization of the electron-hole interaction and self-energy effects induced by the the metallic segment on the electron-hole pair, as well as by the dielectric environment, through an induced charge numerical approach. In neutral matchstick structures the metal NP has only a minor influence ($\sim$1 meV) on the excitonic states. When the  metallic NP is charged the exciton becomes rapidly redshifted and spatially indirect. In contrast, in neutral core-shell structures the exciton energy redshifts by tens of meV.
\end{abstract}


\section{Introduction}

Hybrid nanoparticles -also called nano-hybrids- which combine material segments with different physical and chemical properties at the nanoscale open new venues to applications. For example, while quantum size effects allow to engineer very precisely the electronic properties of semiconductor nanocrystals with a number of possible applications,\cite{Bruchez1998} combining semiconductor nanocrystals with a metallic nanoparticle\cite{Mokari2004,Costi2010} may result in a multi-functional nanomaterial\cite{Wang2009}, with the metallic segment acting, e.g., as an electrical connection\cite{Sheldon2009} or a preferred anchoring site for assembling of complex networks.\cite{Barick2010,Figuerola2009}

Nano-hybrids made from a semiconductor nanorod (SNR) with tips covered by a metal nanoparticle (MNP),\cite{Costi2010} are one example of a growing set of nano-systems with optical properties which are not just the linear combination of those of the constituent segments.\cite{Zhang2006,Artuso2010} A most important property of metal-semiconductor nano-hybrids, for example, is their potential for photo-catalysis. In CdSe-Au and CdS-Au samples the Fermi energy of Au is located in the semiconductor gap. After illumination above the semiconductor gap, the electron-hole pair separates, and fast electron migration to the metallic sector takes place.\cite{Costi2008,Costi2009} As a result, the band gap recombination of the electron-hole pairs is almost suppressed, while the charged metallic tips are able to reduce a molecular acceptor in the solvent. Moreover, electrons can accumulate on the MNP until the Fermi energy aligns with the semiconductor gap, the number of accumulated charges being in the range of few units up to tens of electrons, depending on the dielectric constant of the solvent.\cite{Costi2008,Costi2009} Such systems, based on the direct gap materials CdS and CdSe, absorbing in a wide part of the solar spectrum, are candidates for efficient photo-catalysis, e.g., for hydrogen production. The mechanism for efficient charge migration has not been clarified, though.

\begin{figure}
  \includegraphics[width=10cm]{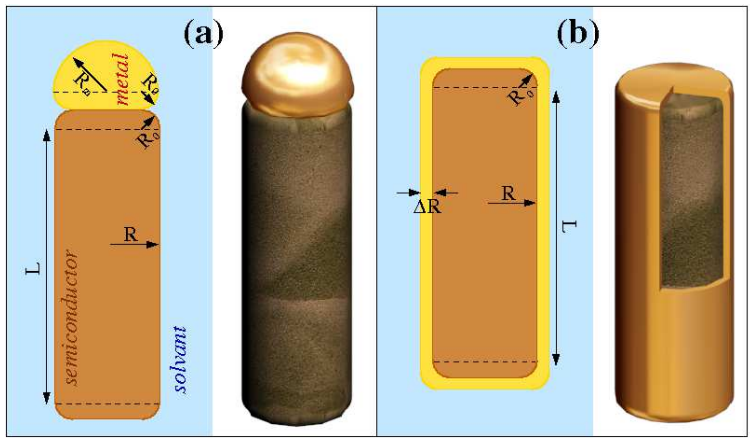}
  \caption{Schematic cross section and 3D geometry of (a) a matchstick and (b) a core-shell metal-semiconductor nano-hybrid.}
  \label{fig:geometry}
\end{figure}

In an attempt to characterize the optical response of metal-semiconductor nano-hybrids, we shall discuss the nature of the excitonic states in systems with different geometries (see \ref{fig:geometry}), namely, a SNR with a MNP attached at one end, also called ``matchstick'' structure, and a SNR covered by a metallic shell (core-shell structure). We shall consider in particular situations in which the excitonic absorption taking place in the SNR segment is well separated with respect to the plasmonic absorption of the MNP or shell. This is the case of CdS-based structures, where the excitonic resonance can be clearly distinguished in absorption spectra\cite{Saunders2006,note-CdSE}. Interestingly, despite the exciton is uncoupled to the plasmon in CdS-Au samples, the exitonic absorption is still found to be blueshifted by few tens of nm in a matchstick structure with respect to bare CdS SNRs, in ensemble measurements.\cite{Saunders2006}

Several energy terms contribute to the excitonic states in the nano-hybrids considered here. In a SNR the excitonic energy is primarily determined by radial quantum confinement effects. The longitudinal confinement, in turn, can be tailored continuously from 0D to 1D regimes.\cite{Saunders2008b}
 In nano-hybrids, in addition to quantum confinement, carries also feel the self-polarization potential (image potential) induced at the semiconductor-metal interface. This is an attractive term for both electrons and holes in the region right outside the metallic surface. For example, in a MNP this may induce bound image potential states.\cite{Rinke2004} In a metal-semiconductor nano-hybrid, where a SNR and a metal are in contact, image potential and lateral confinement may compete, possibly affecting carrier localization.\cite{Demchenko2007}
Due to size quantization, the $e-h$ interaction of photo-excited pairs is also expected to give a large contribution to the spectral properties of SNRs. Clearly, Coulomb interactions will be particularly effective in long SNR where the longitudinal confinement is weak. Furthermore, the overall interaction is strongly enhanced by the dielectric confinement induced by the low-dielectric constant environment, leading, e.g., to strong correlation effects\cite{Climente2009,Royo2010}. The MNP induces a further renormalization of the $e-h$ interaction in the opposite direction, tending to screen the interaction. Lastly, charging of the metal NPs after photoexcitation\cite{Costi2009} will create an internal electric field which may strongly polarize the exciton pair, since it acts in the long, weakly confined direction of the SNR.

In this paper we propose a modelization of excitonic states in metal-semiconductor nano-hybrids, focusing on CdS-metal systems
with the matchstick and the core-shell geometries.
Our modelization takes into account the electron-hole interaction exactly using the configuration-interaction method, with the semiconductor states within an envelope function. Electron and hole states at the excitonic resonance are assumed to be well confined in the SNR,\cite{dePaiva2010} but the influence of the metal Coulomb screening on the exciton carriers is fully taken into account by a screening model.
We will study, in particular, the relative contributions of the different energy terms discussed above, and the shift and nature of the excitonic resonance with respect to charging of the MNP.

\section{Method}

Our aim is to describe the excitonic complex in a SNR which is in close contact with a small MNP or a thin metallic shell, and surrounded by a dielectric environment.
We assume that the plasmon resonance is well separated in energy from the excitonic one, and describe the influence of the metallic segment by a screening model. In other words, we calculate how the excitonic resonance is affected by the polarization of the metal due to an electron-hole pair excitation, neglecting the change in the polarization properties of the metal due to exciton population, which is reasonable in the linear regime.\cite{Zhang2006}
The full Hamiltonian of the exciton in the nano-hybrid reads:


\begin{equation}
	H_X^{NH} = H_{e,h}^{SNR} + H_{e,h}^{pol} + H_C^{SNR} + H_C^{pol} + V_{e,h}^{imp},
	\label{eq:Hfull}
\end{equation}
where
\begin{itemize}

  \item  $H_{e,h}^{SNR}$ is the single-particle Hamiltonian of the electron and hole in the bare SNR,
neglecting the influence of the metal and the environment,

\begin{equation}
	H_{e,h}^{SNR} = H_e^{SNR} + H_h^{SNR}.
\end{equation}

\noindent Here $H_{e(h)}^{SNR}$ is the single-particle Hamiltonian of the photo-excited electron (hole),
which we describe with a 3D single-band envelope function formalism,

\begin{equation}
	H_i^{SNR}(\mathbf{r}_i) =  \frac{\mathbf{p}_i^2}{2 m^*_i} + V_i^{c}(\mathbf{r}_i),
\end{equation}

\noindent with $i=e,h$. $\mathbf{p}$ is the momentum operator, $m^*$ the effective mass and
$V^{c}$ the step-like confinement potential due to the different band alignment of the semiconductor with the metal and the insulating environment.

\item $H_{e,h}^{pol}$ represents the influence of the metal and the environment on the SNR single-particle
states, that is, the self-polarization (image) potentials:

\begin{equation}
	H_{e,h}^{pol} = V^{sp}(\mathbf{r}_e) + V^{sp}(\mathbf{r}_h).
\end{equation}

\noindent Details about the calculation of self-polarization and polarization terms are given in Refs.~\cite{MovillaCPC,BadiaCPC}
See also Supporting Information.

\item $H_C^{SNR}$ is the usual electron-hole Coulomb interaction term for homogeneous dielectrics,
including local screening:

\begin{equation}
	H_C^{SNR}(\mathbf{r}_e,\mathbf{r}_h) = \frac{1}{\varepsilon_s \,\left| \mathbf{r}_e - \mathbf{r}_h \right|},
\label{homo}
\end{equation}

\noindent where $\varepsilon_s$ stands for the semiconductor dielectric constant.

\item $H_C^{pol}$ is the Coulomb attraction between the hole and the electron polarization charges
induced near the interfaces. The term, which will be projected onto a basis of single-particle exciton 
states, can be written as:

\begin{equation}
H_C^{pol}(\mathbf{r}_e^i,\mathbf{r}_h) = \sum_{r,s} \, |r\rangle \, \frac{\rho_{rs}^i(\mathbf{r}_e^i)}{\left| \mathbf{r}_e^i - \mathbf{r}_h \right|} \,\langle s|,
\label{eq:polX}
\end{equation}

\noindent where $|r\rangle$ and $|s\rangle$ are single-electron eigenstates, while
$\rho_{rs}^i(\mathbf{r}_e)$ is the charge density induced by the electron density
charge $|r\rangle |s\rangle$ at the position $\mathbf{r}_e$. See Supporting
Information for further details.

\item $V_{e,h}^{imp}$ is the single-particle potential exerted by $N$ electron
charges trapped in the MNP on the photo-excited electrons and holes.
We model trapped electrons as a multiply-charged impurity
located at the center of the MNP. Then,
$V_{e,h}^{imp}=V_{imp}(\mathbf{r}_e) - V_{imp}(\mathbf{r}_h)$, with:

\begin{equation}
V_{imp} (\mathbf{r}) = \frac{N}{\varepsilon_m \,|\mathbf{r}_0 - \mathbf{r}|} + V_{imp}^{pol}(\mathbf{r}).
\label{eq:imp}
\end{equation}

\noindent Here $\mathbf{r}_0$ is the position of the impurity and $\varepsilon_m$ the dielectric constant of the metal.
$V_{imp}^{pol}$ is the Coulomb interaction with the polarization charges induced by the impurity on the metal surface,

\begin{equation}
V_{imp}^{pol} (\mathbf{r}) = \int \frac{\rho_N^i(\mathbf{r}_0^i)}{|\mathbf{r}_0^i - \mathbf{r}|}\,d\mathbf{r}_0^i.
\label{eq:impol}
\end{equation}

In an ideal metal $\varepsilon_m=\infty$, so the first term in Eq.~(\ref{eq:imp})
is zero and all the interaction takes place through the surface charges.
However, owing to numerical limitations, in our calculations we shall consider a
quasi-metal with large but finite $\varepsilon_m$ (see Supporting Information). 
Yet, the dominant contribution is by far that of $V_{imp}^{pol}$, so that we capture the
essential features of the MNP screening.
\end{itemize}

Note that we have assumed the exciton feels static screening from the metal.
This is an accurate description for nano-hybrids where the excitonic resonance 
does not overlap with interband transitions of the metal, which is the case for
most metals and many nano-hybrids of interest.\cite{Costi2010,Habas2008,Berr2010} 
For CdS-Au nano-hybrids, however, gold interband transitions are already present at 
the frequency of the excitonic resonance,\cite{Saunders2006} which has an impact
on the dielectric response of the metal.\cite{Meyer2006}
In this case, a static screening model implies assuming that the metal is
relaxed when we probe the exciton. This is likely to be the case for photoluminescence 
spectroscopy, but should be taken with caution when comparing with absorption measurements.
Also, for CdS-Au the spectral overlap between semiconductor and metal excitations 
implies F\"orster energy transfer may be important in describing the exciton dynamics.\cite{Govorov2007}
Indeed, this mechanism has been proposed to explain the photoluminescence quenching observed in 
CdSe-Au nano-hybrids.\cite{Costi2010} In CdS-Au systems the influence is less clear because 
F\"orster processes are weaker when excitons and surface plasmons are decoupled.\cite{Govorov2007}

Starting from a cylindrical SNR with radius $R$ and length $L$, below we shall study:
i) matchstick structures (\ref{fig:geometry}(a)) formed by attaching at one end of a SNR a metal hemisphere of radius $R_m$,
and ii) core-shell structure (\ref{fig:geometry}(b)) formed by coating the semiconductor by a metal shell of thickness $\Delta R$.
Both the SNR and the metallic sectors have rounded corners of radius $R_0$. A finite curvature is needed to avoid numerical instabilities and is consistent with TEM images.\cite{Saunders2006} Note that the exact shape of the MNP is not known. Our results do not depend sensitively on the exact choice of the MNP geometry, unless otherwise stated.

The nano-hybrids are supposed to be surrounded by an insulating environment extending to infinity.
Typical values for the confinement potential exerted by the dielectric environment are in the few eV range.
The metallic layer has a Fermi energy which lies in the gap of the semiconductor and has a continuum of states
at the semiconductor conduction edge. Nevertheless, in general the semiconductor states at the gap do not hybridize easily with the metallic continuum, and are well confined in the semiconductor.\cite{Arita2001} For the specific Au/CdS interface it has been shown in Ref.~\cite{dePaiva2010}
that the energy bands near the gap are very little affected by Au adsorption on CdS.
In order to incorporate this effect in the envelope function approximation at the single band level,
we include in our model a large barrier for the semiconductor states at the semiconductor-metal interface.
For simplicity, we use the same barrier height as for the semiconductor-environment interface.
Although the barrier height and precise position are to some extent arbitrary, we have checked that moderate variations
do not affect our results qualitatively. To model the carriers in the semiconductor, we use CdS material parameters
(see Supporting Information).

\section{Results and discussion}

\subsection{Matchstick structure}

CdS nanorods can be grown with ultra-narrow diameter and relatively narrow distribution, from a few units to below the Bohr radius,\cite{Saunders2008,Zeiri2007,Saunders2006} and high aspect ratio (from tens to hundreds nm). Here, we take $R=2.5$ nm and selected lengths in the range $L=5-20$ nm. In this case the diameter slightly exceeds the Bohr radius. We assume the MNP radius is $R_m=2.5$ nm.

\subsubsection{Neutral MNP}

\begin{figure}
  \includegraphics[width=12cm]{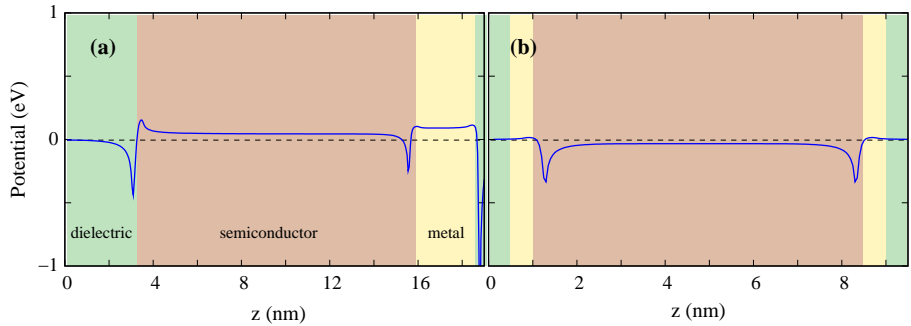}
  \caption{Self-polarization potential along the longitudinal axis of (a) a matchstick structure with $L=12.5$ nm and (b) a core-shell structure with $L=7.5$ nm.
	   The dashed line highlights the zero potential position. Background colors indicate the
	   type of material in each region: semiconductor (brown), metal (yellow) or dielectric (green).}
  \label{fig:self}
\end{figure}

\begin{figure}
  \includegraphics[width=5cm]{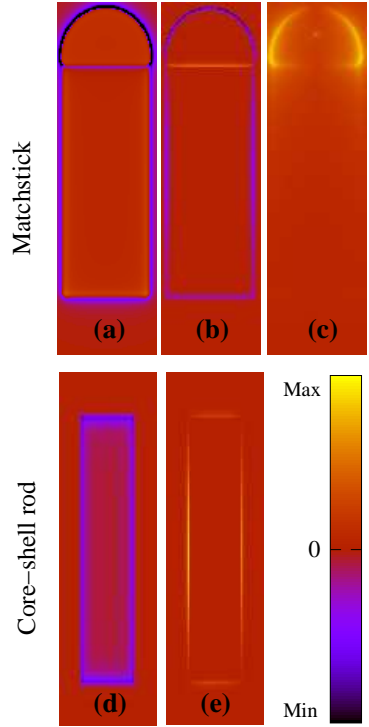}
  \caption{Contour plots of different terms arising from dielectric mismatch in a matchstick (upper row) and a core-shell structure (lower row).
	   (a) and (d): self-polarization potential, $V^{sp}$. (b) and (e): polarization charge induced by the dominant electron configuration,
	   $\rho_{e0}^i$.
	   (c): Coulomb potential arising from $N=2$ electrons localized in the center of the MNP, $V_{imp}$.}
  \label{fig:terms}
\end{figure}

We start  by investigating the different energy contributions with a neutral MNP. In this case, $V_{e,h}^{imp}=0$ and the Hamiltonian reads:

\begin{equation}
	H_X^{NH} = H_{e,h} + H_C,
	\label{eq:H_N0}
\end{equation}

\noindent where $H_{e,h}=H_{e,h}^{SNR} + H_{e,h}^{pol}$ are the single-particle terms
and $H_C=H_C^{SNR} + H_C^{pol}$ the electron-hole interaction.
The complex dielectric configuration influences the excitonic resonance via the image potential
$H_{e,h}^{pol}$ and the Coulomb polarization term, $H_{C}^{pol}$.
To understand the role of these terms, in \ref{fig:self}(a) we show the calculated image potential
along the longitudinal axis of the structure.
Carriers are confined inside the SNR, whose dielectric constant is intermediate between that of the dielectric environment
and that of the metallic tip. Therefore, the potential $V^{sp}$ felt by photo-excited electrons and holes has
(i) a repulsive character ($V^{sp}>0$) in most of the semiconductor region,
owing to the influence of the dielectric environment in the radial direction,
(ii) a potential barrier at the interface with the dielectric environment ($z\sim 4$ nm),
and (iii) a potential well near the interface with the MNP ($z\sim 16$ nm).
A more complete perspective of $V^{sp}$ is given in \ref{fig:terms}(a), which shows the
corresponding 2D plot.

Next we consider the Coulomb polarization terms, $H_C^{pol}$.
For a preliminary insight into its influence, \ref{fig:terms}(b) shows the polarization
charges induced by the electron in the lowest-energy orbital ($|r\rangle=|s\rangle=|0\rangle$),
 $\rho_{e0}^i$, which is the dominant term in Eq.~(\ref{eq:polX}).
One can see that the charges distribute near the interfaces,
and they have different sign on the semiconductor interfaces with the environment
(where they are negative) and metal (where they are positive). Thus, Coulomb interaction
with the hole will be enhanced (weakened) near the dielectric environment (conducting metal).

\begin{figure}
  \includegraphics[width=7.5cm]{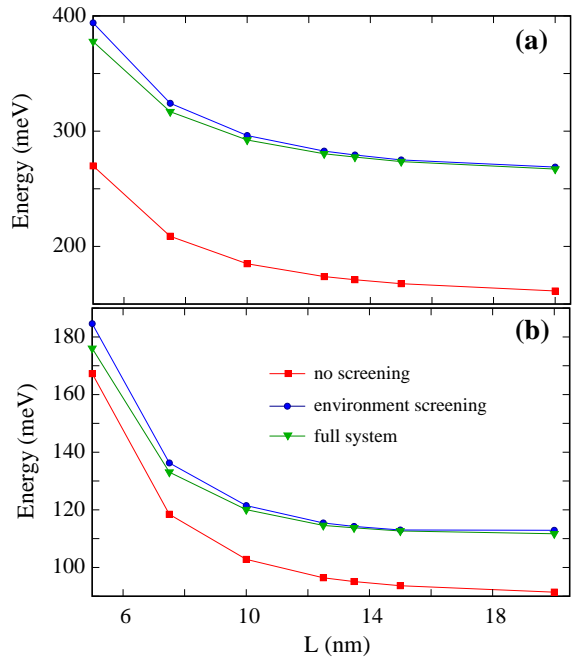}
  \caption{Energy contributions to the excitonic energy of a matchstick structure. (a) single-particle energy of the electron-hole pair, neglecting dielectric confinement effects (red squares), including screening from the environment (blue circles) and that from environment plus metal (green triangles). (b) same but for excitons, including Coulomb interaction terms.}
  \label{fig:energy-contributions}
\end{figure}

We are now in the position to perform quantitative estimates of the exciton energy
in matchstick structures. \ref{fig:energy-contributions}(a) shows the single-particle contribution to the energies of an electron-hole pair as a function of the SNR length. To highlight the different contributions, we compare a bare SNR system (no dielectric mismatch effects, red squares) with a SNR in a dielectric environment (blue circles)
and with the attached MNP (green triangles). In all cases the confinement energy decreases with increasing $L$ due to the relaxation of the longitudinal confinement.\cite{HuSCI,Royo2010}
In addition, the dielectric environment is responsible for a large, rigid blueshift
of about 100 meV, independent of $L$. This is due to the repulsive nature of $V^{sp}$,
mostly arising from the radial dielectric confinement induced by the insulating environment. The inclusion of the metal has the opposite effect, namely to redshift the electron-hole pair energy,
due to the attractive potential near the metal interface. However, this contribution
is only significant for short $L$. For structures with typical length of $L\sim15$ nm, it is
about 1 meV.

A similar trend holds for the exciton energy, i.e., including Coulomb interaction terms, as shown in \ref{fig:energy-contributions}(b).
There are however two visible differences. First, the magnitude of the blueshift due to the dielectric environment is now reduced to
less than $20$ meV. This is because $H_C^{pol}$ shifts the energy in opposite direction with respect to $V^{sp}$,\cite{LannooPRL,BolcattoPRB}
with the the latter being slightly larger than the former\cite{Royo2010}, which leads to a partial cancellation.
Second, the blueshift decreases for short SNR. This is because Coulomb interaction terms, including $H_C^{pol}$,
are enhanced in strongly confined structures, while $V^{sp}$ remains as a short-ranged interaction near the interfaces.

One important point concerns the nature of the excitonic transition. In particular, one may wonder whether the excitonic wavefunction is distorted or even localized by the presence of the metallic segment and the ensuing image potential well. To this end, we calculate the
average electron and hole charge density, taking the charge density
of the exciton ground state and integrating over the other particle's
degrees of freedom. For example, the electron charge density in the excitonic ground state is
evaluated as
\begin{equation}
    \rho_e(\mathbf{r}_e)=\int \Psi_X(\mathbf{r}_e,\mathbf{r}_h)^* \;\Psi_X(\mathbf{r}_e,\mathbf{r}_h)\, d\mathbf{r}_h
\end{equation}
where $\Psi_{X}(\mathbf{r}_e,\mathbf{r}_h)$ is the lowest-energy eigenstate of Eq.~(\ref{eq:Hfull}).
An example is shown in \ref{fig:wavefunctions} ($N=0$) for a structure of intermediate length. Both the electron and hole are localized in the middle of the SNR, as would be the case in the absence of the MNP. No sizable distortion is observed, despite the asymmetry of the structure. As we shall also discuss for core-shell structures, the image potential is usually too short ranged to localize carriers in CdS. However, this also depends on the effective mass and the quantum confinement. For specific materials, such as GaP,\cite{BadiaCPC} and/or crystallographic direction, therefore, one might obtain interface localized excitons.

\subsubsection{Charged MNP}

Next we investigate the influence of charging the MNP, which might be the result of fast electron migration to the metallic sector and hole scavenging.\cite{Costi2008,Costi2009} Upon continuous optical excitation, a new exciton is formed in the presence of the electron(s) trapped in the MNP.
The Hamiltonian is then given by:

\begin{equation}
	H_X^{NH} = H_{e,h} + V_{e,h}^{imp} + H_C.
	\label{eq:H_N}
\end{equation}

\begin{figure}
  \includegraphics[width=12cm]{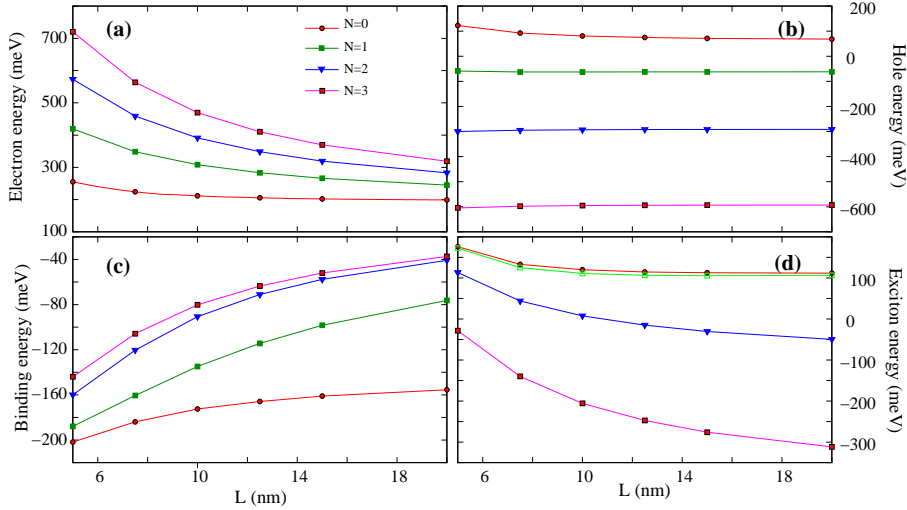}
  \caption{Influence of $N$ electrons trapped in the MNP on the energy of carriers confined in the SNR. (a) electron energy.
	   (b) hole energy. (c) exciton binding energy. (d) exciton energy.}
  \label{fig:energy-charging}
\end{figure}

The electrons trapped inside the MNP redistribute over the metal surface yielding a
potential $V_{imp}$, as plotted in \ref{fig:terms}(c). The effect of this additional
term is shown in \ref{fig:energy-charging}, with the MNP tip charged with up to three electrons.
The figure shows the (a) electron confinement energy, $E_e$; (b) hole confinement energy, $E_h$; (c) electron-hole binding energy, $E_{b}$,
and (d) exciton energy $E_X=E_e+E_h+E_{b}$.

Being $V_{imp}$ a repulsive term for electrons, it shifts upwards the electron energy, $E_e$, as shown in \ref{fig:energy-charging}(a). The repulsion is weaker for the larger $L$ because the electron localizes on the far side of the SNR to minimize the repulsions with the metal. As a result, $E_e$ is more sensitive to $L$ in the presence of a charged MNP than with neutral MNP (compare, e.g., $N=3$ and $N=0$ curves in panel (a)).

For holes, see \ref{fig:energy-charging}(b), $V_{imp}$ is attractive. Accordingly, the confinement energy, $E_h$, becomes negative upon charging, and the $L$ dependence almost disappears because holes become trapped at the semiconductor-metal interface, which clearly dominates over the weak longitudinal confinement.

\ref{fig:energy-charging}(c) shows the electron-hole binding energy, $E_b$. Its absolute value decreases with $N$, since electron-hole pair is pulled apart by the electric field induced by the charged MNP, reducing their mutual Coulomb interaction. An interesting point is that this effect tends to saturate at $N=2$, since at this value the electron-hole pair is already as splitted as the longitudinal confinement allows. An additional charge ($N=3$) does not substantially change the binding energy for the long structures. The $N$-dependence remains efficient only for the shorter structures, where the quantum confinement forces the electron and hole to share the same spatial region.

The exciton energy, $E_X$, see \ref{fig:energy-charging}(d), turns out to be dominated by the hole attraction due to the charged MNP. As a result, the exciton energy decreases with $N$, in a similar fashion to holes. Yet, excitons are more sensitive than holes to the rod length, which is a consequence of the significant variations of $E_e$ and $E_{b}$ with $L$.
For long structures, the shift is a large fraction of an eV, and the resonance may fall in the broad plasmon resonance.\cite{Saunders2006} 

\begin{figure}
  \includegraphics[width=12cm]{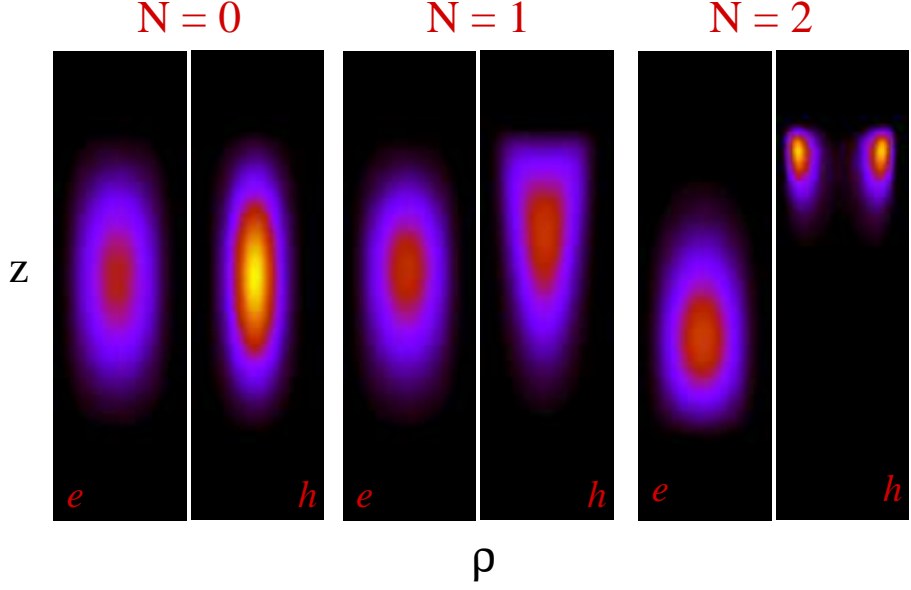}
  \caption{Localization of the electron ($e$) and hole ($h$) of an excitonic pair in a matchstick with $L=12.5$ nm
as a function of the number of electrons trapped in the MNP.}
  \label{fig:wavefunctions}
\end{figure}

The charge densities as a function of the number of
electrons in the MNP are plotted in \ref{fig:wavefunctions}. For each $N$, left (right) panels show the electron (hole) charge density. It can be clearly seen that an increasing electron charging pushes electrons and holes in opposite directions, despite the reciprocal Coulomb attraction.
A striking feature is that, for $N \ge 2$, holes do not simply localize over the entire semiconductor/metal interface.
Instead, they show localization in a ring-like state. This result can be understood from the shape of $V_{imp}$.
As can be seen in \ref{fig:terms}(c), $V_{imp}$ is stronger
on the metal/dielectric interface than on the metal/semiconductor
one, because of the larger dielectric mismatch.
Besides, it is stronger near the corners of the hemisphere owing to the larger curvature,
which translates into a larger gradient of dielectric constants.
This favors hole localization near the corners of the SNR.\cite{shapeMNP}
 Note that holes are particularly sensitive to dielectric mismatch effects
(an attractive potential in this case) in the radial direction because
of the larger in-plane mass, $m^*_h(\perp)$, as compared to that along the
growth direction $m^*_h(z)$.

\subsection{Core-shell nanorods}

In this section we study the effect of the dielectric mismatch in a
SNR with $R=2$ nm, $L=7.5$ nm, coated with a metal shell of
thickness $\Delta R=0.5$ nm and embedded in a dielectric environment.

 \ref{fig:self}(b) shows the longitudinal profile of the image potential.
One can immediately spot important differences as compared to the case of the matchstick structures, namely, (i) $V^{sp}$ is now attractive in the core, and (ii) there are only potential wells near
the semiconductor interfaces. The differences are due to the dominant contribution
of the metal shell all over the SNR, as opposed to the matchstick case, where most regions felt
a dominant contribution from the dielectric environment. The radial profile of $V^{sp}$ is
similar to the longitudinal one, as can be seen in the 2D plot of \ref{fig:terms}(d).
The polarization charges induced by the electron in the lowest-energy orbital, $\rho_{e0}^i$,
are illustrated in \ref{fig:terms}(e).
By comparing with the results for corresponding matchstick structure,
panel (b), one can also notice qualitative differences. Namely, the charges are
now repulsive for the hole all over the semiconductor surface.

\begin{figure}
  \includegraphics[width=12cm]{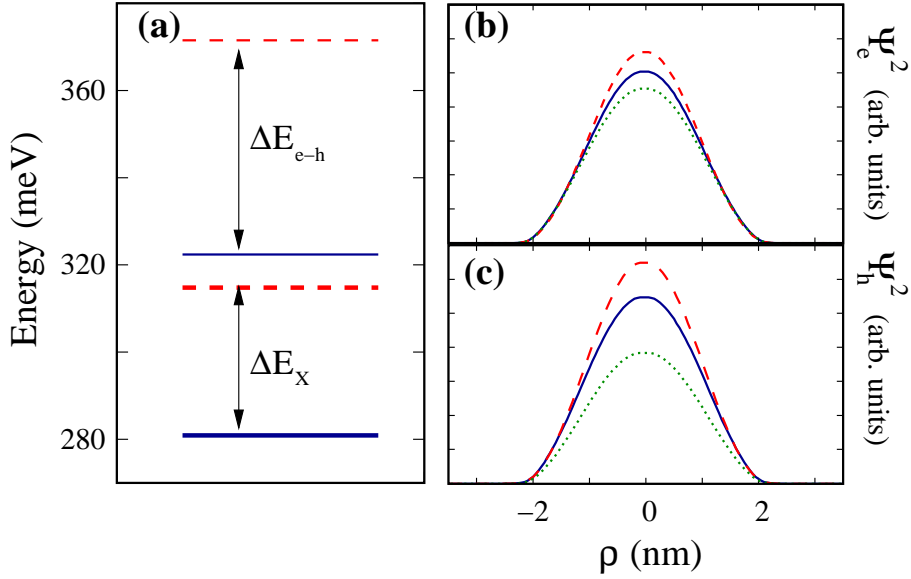}
  \caption{Influence of the dielectric confinement in a core-shell nano-hybrid.
	   (a) single-particle energy of the electron-hole pair (thin lines) and excitonic energy (thick lines).
	   (b) and (c): radial cross-section of the electron and hole charge density in the excitonic complex, respectively.
	   Solid and dashed lines include (exclude) dielectric mismatch terms. In (b) and (c), dotted lines are
	   used to represent single-particle densities.}
  \label{fig:coreshell}
\end{figure}

The effect of the polarization terms on the energy levels is studied in
\ref{fig:coreshell}(a). Solid (dashed) lines are the results including
(excluding) such terms. For the single-particle electron-hole pair
(thin lines), the image potential translates into a redshift
of $\Delta E_{e,h}=-56.7$ meV. For the interacting exciton (thick lines),
this is partly quenched by the polarization charges weakening Coulomb
interaction, so that the redshift reduces to $\Delta E_{X}=-41.5$ meV.

$V^{sp}$ in \ref{fig:terms}(d) suggests that carriers in core-shell structures may
localize in the potential well near the metal surface.
The large radial mass of holes in CdS, $m^*(\perp)=2.56\,m_0$,
would further support this phenomenon because it entails moderate kinetic energy
in spite of the strong radial confinement.
The situation may however change for excitons, because $H_C^{pol}$
is expected to counteract $V^{sp}$.
To investigate this point, we have calculated the charge densities of
the electron and hole within the exciton complex and plotted them in
\ref{fig:coreshell}(b) and (c), respectively.
Solid (dashed) lines include (exclude) polarization terms.
One can see that the inclusion of such terms visibly reduces the density
in the center of the rod, but no drastic localization near the metal occurs.
As a matter of fact, the single-particle states (dotted lines) do not localize either.
This shows that the image potential well formed by $V^{sp}$ is generally too thin to compete
against quantum confinement.

\section{Conclusion}

We have developed a model to study excitonic resonances in semiconductor-metal nano-hybrids, whereby the excitonic states are subject to the screening of nearby metallic segments and
dielectric environments. We have explicitly estimated the influence of the different energy contributions to the excitonic frequency in the case of CdS-based nano-hybrid with the matchstick and the core-shell geometry.

In neutral matchstick structures, the metal sector has a minor influence on the energy and charge density localization. 
This suggests that the blueshift observed in related experiments upon Au absorption\cite{Saunders2006} does not arise 
from polarization effects, rather from structural changes or by interactions with the interband transitions of the metal.
The influence of the metal becomes however critical when the MNP is charged. The exciton becomes rapidly redshifted
with increasing number of charges in the metal. Moreover, the Coulomb interaction with the additional charges overcomes the electron-hole interaction, and the exciton becomes spatially indirect.
The sensitivity of the excitonic resonance upon charging of the MNP suggests the use of matchsticks for sensing of charged molecules.

In core-shell nano-hybrid, on the contrary, the uniform metal coating redshifts the
exciton energy by tens of meV. The stronger influence of the metal
as compared to the matchstick is due to dominant role of the radial confinement.
For certain nano-hybrid geometries, we predict the existence of exciton
states where the hole is attached to the lateral sides of the SNR.

\section{Associated Content}

{\bf Supporting Information.} Details on the method of induced charges computation.
Calculation procedure and parameters used for solving Eq.~(\ref{eq:Hfull}). 
This material is available free of charge via the internet at
http://pubs.acs.org.

\acknowledgement

We gratefully acknowledge useful discussions with F. Valle\'e, N. Del Fatti, U. Banin,
C. S\"onnichsen, R. di Felice and U. Hohenester.
We acknowledge partial financial support from EU through the NanoSci ERA-Net scheme, program "single-nanoHybrid", MCINN project CTQ2008-03344,
MIUR-FIRB Italia-Canada RBIN06JB4C, UJI-Bancaixa project P1-1A2009-03, the Ramon y Cajal program (J.I.C.),
CNR-Short mobility 2010, and UJI-Bancaixa INV-2009-11.
Calculations have been performed under the CINECA-Iniziativa calcolo parallelo 2010.

\section{Supporting Information}

\subsection{Induced charge computation method}

Here we summarize the theory of the screening model we use. 
We start from the Poisson equation for inhomogeneous dielectric media,

\begin{equation}
\nabla \left[ \varepsilon(\mathbf{r}) \, \nabla \psi(\mathbf{r}) \right] = - 4 \pi\,\rho(\mathbf{r}),
\label{poisson}
\end{equation}

\noindent where $\varepsilon(\mathbf{r})$ is the position-dependent dielectric constant,
$\psi(\mathbf{r})$ the electrostatic potential and $\rho(\mathbf{r})$ the source charge
(electron, impurity, etc). We want to replace Eq.~(\ref{poisson}) by the Poisson equation
in the vacuum of the source charge plus an additional induced charge, $\rho_i(\mathbf{r})$, 
which implicitly accounts for local screening and interface effects,

\begin{equation}
\nabla^2  \psi(\mathbf{r}) = - 4 \pi\, (\rho(\mathbf{r})+\rho_i(\mathbf{r})).
\label{poisson2}
\end{equation}

\noindent Taking Eq.~(\ref{poisson2}) into Eq.~(\ref{poisson}) one obtains:

\begin{equation}
\rho_i(\mathbf{r}) = \rho\,(\frac{1}{\varepsilon(\mathbf{r})} - 1) + \frac{\nabla \varepsilon(\mathbf{r}) \, \nabla \psi(\mathbf{r})}{4\pi\,\varepsilon(\mathbf{r})}.
\label{rho_i}
\end{equation}

\noindent The first term of $\rho_i(\mathbf{r})$ is the contact interaction correction, $\rho_i^c$. 
When added to the source charge, it gives the locally screened charge, 
$\rho(\mathbf{r})+\rho_i^c(\mathbf{r})=\rho(\mathbf{r})/\varepsilon(\mathbf{r})$. 
The second term is finite only when $ \nabla \varepsilon(\mathbf{r}) \neq 0$, and it gives rise to the interface
effects. We can then split the electrostatic potential in Eq.~(\ref{poisson2}) as 
$\psi(\mathbf{r})=\psi_s(\mathbf{r})+\psi_p(\mathbf{r})$, so that:

\begin{eqnarray}
\nabla^2  \psi_s(\mathbf{r}) = - 4 \pi\, \frac{\rho(\mathbf{r})}{\varepsilon(\mathbf{r})},\\
\nabla^2  \psi_p(\mathbf{r}) = - 4 \pi\, \tilde{\rho_i}.
\end{eqnarray}

\noindent where $\tilde{\rho_i}=\rho_i - \rho_i^{c}$ is the induced charge excluding the contact charge
(first term of \ref{rho_i}).  $\psi_s$ is the usual homogeneous electrostatic potential, 
as in Eq.~(\ref{homo}), while $\psi_p$ is the polarization potential, as in Eq.~(\ref{eq:polX}).
Further details about the calculation of $\rho_i$ can be found in Ref.\cite{MovillaCPC}.

\subsection{Calculation procedure}
\begin{figure}
  \includegraphics[width=6cm]{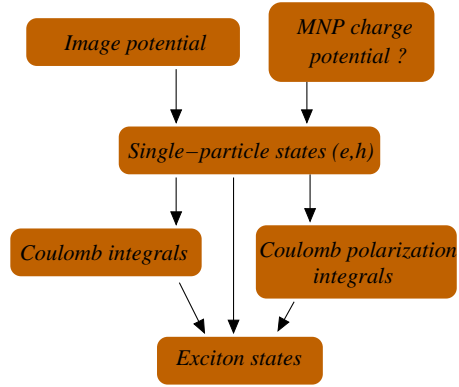}
  \caption{Flow diagram showing the procedure for solving Eq.~(\ref{eq:Hfull}).}
  \label{fig:flow}
\end{figure}

Hamiltonian (\ref{eq:Hfull}) is solved following the scheme of \ref{fig:flow}.

First, knowing the geometry and dielectric constants of the system, 
we calculate the image potential, $V^{sp}$. This is done numerically 
using the induced charge computation method for arbitrary
geometries with cylindrical symmetry.\cite{MovillaCPC}
Solving Eq.~(\ref{rho_i}) for the self-energy requires a 3D discretization of the space,
with a high density of points near the metal interface, where induced charges are essentially
localized on the surface. This leads to huge and dense linear systems $A\,x = b$,
which we solve using a recently developed library combining $L\,U$ decomposition of 
the matrix $A$ with state-of-the-art out-of-core algorithms.\cite{BadiaCPC}
If the MNP is charged, we further compute the potential induced by the electrons
near the metal surface, $V_{imp}^{pol}$, Eq.~(\ref{eq:impol}). 
The induced charge $\rho_N^i(\mathbf{r}_0^i)$ is also calculated using the induced 
charge computation method.

Second, we determine the single-particle electron and hole states.
To this end, we integrate $H_{e,h}^{SNR} + H_{e,h}^{pol} + H_{e,h}^{imp}$
using cylindrical coordinates and finite differences. 
We then build a basis set to expand the exciton
states as Hartree products between electron and hole states.
The basis includes 20 low-energy electron spin-orbitals 
(12 with azimuthal angular momentum $m_z=0$ and 4 with $m_z=\pm 1$),
and 40 low-energy hole spin-orbitals (24 with $m_z=0$ and 8 with $m_z=\pm 1$).
Next, the Coulomb interaction is taken into account, $H_C^{SNR} + H_C^{pol}$.
$\rho_{rs}^i$ is obtained following Refs.~\cite{MovillaCPC,BadiaCPC},
and the Coulomb matrix elements are evaluated using Monte-Carlo routines.
Finally, the Hamiltonian matrix including single-particle states and Coulomb 
matrix elements is diagonalized using a configuration interaction method,\cite{citool,citooldoc,RajadellJPCc} 
which yields the exciton states.

\subsection{Parameters}
The same material parameters are used for both matchstick and core-shell structures.
We use CdS effective masses. For electrons we take $m^*_e=0.21\,m_0$,
where $m_0$ is the free electron mass.\cite{Madelung} The hole effective mass in wurtzite structures
is strongly anisotropic. We take $m^*_h(z)=0.22\,m_0$ ($m^*_h(\perp)=2.56\,m_0$) for the longitudinal
 (radial) direction. These values correspond to the mass of $C$-band holes, which is the
dominant component of low-energy holes in CdS nanorods.\cite{PlanellesJPCc}
$V^{c}$ is zero inside the semiconductor and $4$ eV outside (the electroaffinity of the semiconductor).
The dielectric constant is $\varepsilon_s=9.2$ in CdS,\cite{Madelung}
$\varepsilon_e=3.0$ in the dielectric environment and $\varepsilon_m=100.0$ in the metal.
Increasing further the dielectric constant of the metal has only a minor quantitative
effect on the results, at the price of a fast increase of the computational burden.
A rapid but continuous variation of $\varepsilon$ in a region of $5$ \AA~around the interfaces
is assumed to avoid singularities in the self-polarization potential.\cite{BolcattoJPCM}
In all cases we take a small corner radius, $R_0=0.25$ nm (see \ref{fig:geometry}).




\bibliography{paper}

\begin{mcitethebibliography}{39}
\bibitem[sha()]{shapeMNP}
 Radial localization is attenuated as the MNP becomes spherical.\relax
\mciteBstWouldAddEndPunctfalse
\mciteSetBstMidEndSepPunct{\mcitedefaultmidpunct}
{}{\mcitedefaultseppunct}\relax
\EndOfBibitem
\end{mcitethebibliography}

\end{document}